\documentclass[useAMS,usenatbib, usegraphicx]{mn2e}

\usepackage{epsfig}
\include{epsf}

\def\apj{{\rm ApJ}}
\def\apjs{{\rm ApJS}}
\def\mnras{{\rm MNRAS}}

\def\etal{{\rm et~al.\ }}
\def\hmpc{\;h^{-1}{\rm Mpc}}

\def\kms{{\rm \;km\;s^{-1}}}

\def\kmsmpc{\kms\;{\rm Mpc}^{-1}}
\def\msun{{\rm M_{\odot}}}
\def\hmsun{\;h^{-1}{\rm M_{\odot}}}
\def\lya{{Ly$\alpha$} }

\def\taueff{$\tau_{\rm eff}$ }

\def\simlt{\lower.5ex\hbox{$\; \buildrel < \over \sim \;$}}
\def\simgt{\lower.5ex\hbox{$\; \buildrel > \over \sim \;$}}

\title[Constraining quasar host masses with the \lya\ forest]
{Constraining quasar host halo masses with the strength of nearby
  Lyman-alpha forest absorption
}
\author[Y.-R. Kim and R. A. C. Croft]{Young-Rae Kim$^{1}$\thanks{E-mail:
    yr@cmu.edu} and Rupert A. C. Croft$^{1}$\\
$^{1}$Physics Department, 
Carnegie Mellon University, Pittsburgh, PA 15213, USA}

\begin{document}

\date{\today}

\pagerange{\pageref{firstpage}--\pageref{lastpage}} \pubyear{2005}

\maketitle
\label{firstpage}

\begin{abstract}
Using cosmological hydrodynamic simulations we measure the mean 
transmitted flux in the \lya\ forest for quasar sightlines
that pass near a foreground quasar. We find that the trend of absorption with
pixel-quasar separation distance can be fitted using a simple power law form
including the usual correlation function parameters $r_{0}$ and $\gamma$ 
so that ($\left<F(r)\right> = \sum
\exp(-\tau_{\rm eff} (1+(\frac{r}{r_{0}})^{-\gamma}))$).
From  the simulations we find the relation
between $r_{0}$ and quasar mass 
and formulate this as a way to estimate quasar host dark matter halo masses,
quantifying uncertainties due to cosmological and IGM parameters,
 and redshift errors. 
With this method, we  examine data for  $\sim3000$  quasars
from the Sloan Digital Sky Survey (SDSS) Data Release 3, assuming that
the effect of ionizing radiation from quasars (the so-called
transverse proximity effect) is unimportant (no evidence for it
is seen in the data.)
We find that the best fit host halo mass for SDSS quasars
with mean redshift $z=3$ and absolute G band magnitude $-27.5$ is
 $\log_{10} {\rm M/\msun}$=$12.48^{+0.53}_{-0.89}$. We also use the
Lyman-Break Galaxy (LBG) and \lya\ forest data of Adelberger et al
 in a similar fashion to
constrain the halo mass of LBGs to be 
$\log_{10} {\rm M/\msun}$=$11.13^{+0.39}_{-0.55}$,
 a factor of $\sim 20$ lower than the bright quasars. 
In addition, we study the redshift distortions of the
 \lya\ forest around quasars,
 using the simulations. We use the quadrupole to 
monopole ratio of the quasar-\lya\ forest correlation function
 as a measure of the squashing effect. We find that this
does not have a measurable dependence
on halo mass, but may be useful for constraining cosmic geometry.

\end{abstract} 


\section{Introduction} 
The Lyman-alpha forest of absorption lines seen in the spectra of 
quasars see (e.g., Rauch 1998 for a review) can be 
related in theories of structure formation to
fluctuations in the matter distribution. Because the fluctuations are
only weakly non-linear, the photoionized gas where they arise traces the
dark matter distribution that dominates the gravitational potential 
very well, and the \lya\ forest can be used to give us information on 
structure in the dark matter (e.g., Cen \etal 1994; Zhang, Anninos,
 \& Norman 1995;
Petitjean, M\"ucket, \& Kates 1995; Hernquist \etal 1996; Katz \etal
1996; Wadsley \& Bond 1997; Theuns \etal 1998; Dav\'e et al. 1999).
 We can therefore use the \lya\
forest absorption around galaxies and quasars to tell us about the 
profile of dark matter around these objects. The clustering
of dark matter around dark matter halos is expected to be related
to their mass. This has been used by e.g., Seljak \etal (2005) and 
Mandelbaum \etal (2006), 
to constrain the mass of galaxy halos using the matter
distribution around them inferred from weak lensing observations. In the
present paper, we propose to infer related constraints on the
mass of quasar hosts and high redshift galaxies from the \lya\ forest-derived
mass profiles around them.

Some current mass constraints for quasar hosts and high redshift galaxies
come from considering the amplitudes of their two point-correlation 
functions.  For example, Croom \etal 2005 analyzed the 
autocorrelation function of quasars in 2dF data, and  Myers
\etal (2006) the clustering of 300,000 photometrically 
classified SDSS QSO to find constraints on quasar host masses.

It has been known since the work of 
Bajtlik \etal (1988) that the \lya\ forest in the spectrum of 
quasars shows evidence of decreasing absorption as the quasar's emission
redshift is approached. This is known as the proximity effect, and is believed
to be due to photoionization from the quasar itself adding to the
mean background photoionizing radiation field and decreasing the 
amount of neutral hydrogen in the vicinity of the quasar. Knowing  
quasar observed luminosities, the observed decrease in \lya\ absorption
has been used to estimate the intensity of  the mean cosmic background
photoionizing radiation field (see e.g., Scott \etal 2000). A related
process is the possible effect of other (foreground) quasars on the
\lya\ forest of other quasars whose sightlines pass close by. This 
effect, known as the transverse, or foreground
 proximity effect has not so far been observed,
and in fact there have been several non-detections of the
 effect (e.g. Schirber \& Miralda-Escude 2004, Croft 2004 ).

This could
be explained if for example quasar radiation is beamed, so that the usual
proximity effect would occur but off-axis sightlines would not be affected.
Alternatively, the quasar lifetime could be short enough that light
has not had time to travel transversely to adjacent sightlines (for example if
they are 10$\hmpc$ away, a quasar lifetime $\simlt 10$ Myr would leave
them unaffected. The two studies referred to above point instead to
increased \lya\ absorption in the parts of sightlines that pass close
to foreground quasars. This is expected, as quasars should be hosted by
massive dark matter halos which are in overdense regions. We will use this
fact along with theoretical predictions in the context of the Cold Dark Matter
model to constrain the quasar halo masses. The same can also be done for
high redshift galaxies where the quasar sightlines pass close to
galaxies. Here the proximity effect is known definitively to be too small
to affect the spectra (see e.g. Bruscoli \etal 2003)
and instead increased absorption is also seen (Adelberger \etal 2003).
Some related work on the \lya\ optical depth around quasars has been carried 
out by Rollinde \etal 2005 and the distribution of damped \lya\
absorption around foreground quasars by Hennawi \etal (2006).

Looking at the profile of \lya\ forest absorption around quasars in 
two dimensions, one parallel and one perpendicular to the line of sight
allows one to quantify redshift distortions. As the gravitating mass
governs the amount of squashing seen (e.g., Kaiser 1987, Regos \& Geller
1989), measuring it may allow us to infer the mass enclosed within a given 
radius in a complementary way to looking at the mean strength of
\lya\ absorption.

Our plan for the paper is as follows: In \S2, we describe the
hydrodynamic simulations we used, as well as how \lya\ spectra were made. 
In this section we also compute the mean density and velocity profiles as
a function of halo mass, directly from the simulation mass and 
velocity data. We also compute the mean \lya\ forest flux as a function of
quasar-pixel distance, for different mass bins. In \S3 we describe how
we fit the \lya\ forest flux-distance trend with a simple power-law model and
how the power law parameters depend on the halo mass. In \S4 we describe
the data from the SDSS and the LBG data from Adelberger and use our
results from \S3 to constrain the halo masses. 
In \S5, we examine the redshift space anisotropy
of clustering. In \S6 we discuss our results
and conclude.

\section{Simulated QSO spectra}

\subsection{Simulations}
 
We use two large N-body + hydrodynamics simulations
of a $\Lambda$ CDM cosmology to make our spectra. The two simulations
were run using the code GADGET-2 (Springel, Yoshida \& White
2001, Springel 2005), and are described more fuly in Nagamine \etal 2005.
The simulations have the same cosmological parameters
 ( $\Omega_{\Lambda} = 0.7, \Omega_{m} = 0.3,
\Omega_{b} = 0.04$ and $H_{0}=70 \kmsmpc$, $\sigma_8$=0.9) but
have different box sizes and particle numbers. This means that they
have significantly different mass and spatial resolutions so that we
can use them for a resolution study, in order to make sure that 
our results have converged. Both simulations include gas physics, heating,
 cooling, a prescription for star formation (Springel \& Hernquist 2003)
and the effect of stellar feedback on has properties.

One simulation is the G6 run, a cubic box of 100 
$\hmpc$ on each side, with $486^{3}$ dark matter particles and
$486^{3}$ gas particles. 
This results in an initial mass per gas particle of $9.7 \times 10^{7} \msun$
and dark matter mass $6.3 \times 10^{8} \msun$.
The other is the D5 run, a cubic box of side 
length $33.75 \hmpc$ and $2\times324^{3}$ particles.   

\subsection{Mock spectra}

We use simulation outputs from redshift $3$ to make
mock \lya\ spectra.  This was done in the usual
manner, by integrating through the SPH kernels of the particles to
obtain the neutral hydrogen density field, and then convolving with
the line of sight velocity field (see e.g., Hernquist \etal 1996)
The mean UV background radiation intensity
has been normalized so that the spectra have 
a mean effective optical length $\tau_{\rm eff}=0.4$ (where 
$\tau_{\rm eff}=-\ln \left<F\right>$)
or equivalently a mean flux $\left<F\right>$=0.67.
We make 5000 spectra from each simulation, where the sightlines are parallel
to the box axes and the $x-y$ positions of the sightlines are
picked randomly. 

\subsection{Quasar hosts}
For halo finding, we use the friends-of-friends (fof) method (e.g., Huchra
\& Geller 1982). A  particle is defined to belong to a  group if it is
within  some  linking  length  ($b$)  of any  other  particle  in  the
group. We select clusters  by using  $b=0.2$ where $b$ is
the linking length as a  fraction of the mean particle separation.
This definition of dark matter haloes was shown by Jenkins \etal (2001)
to yield a mass function with a univeral form.
We use these dark matter halos as our quasar hosts in the simulation.
We bin the quasar hosts in terms of their mass. 
For each simulation we use has 5 mass bins, logarithmically spaced. There
are 4 overlapping bins. D5 (G6) has one more bin smaller (larger) than the
other 4. This is because in the D5 simulation there are too few halos with
masses $> 10^{12} \msun$ and in the G6 simulation the mass
per particle is too large to resolve halos with masses $< 10^9 \msun$.

\subsection{Mean baryonic density and infall velocity as
a function of quasar-pixel distance}

The baryonic density around quasars is related
to  the amount of neutral hydrogen
 which absorbs light via the \lya energy transition. 
As we have mentioned in \S1, 
 baryons fall into gravitational wells formed by dark matter,
and so the baryonic distribution will mimic that of dark matter.
With two quasars with close angular positions, but at different
redshifts, we can therefore investigate the distribution of dark
 matter by studying the \lya absorption lines around a foreground quasar.
 Quasars have high 
luminosities despite their high redshift, and so allow us to 
probe the high redshift Universe.

\begin{figure}
\psfig{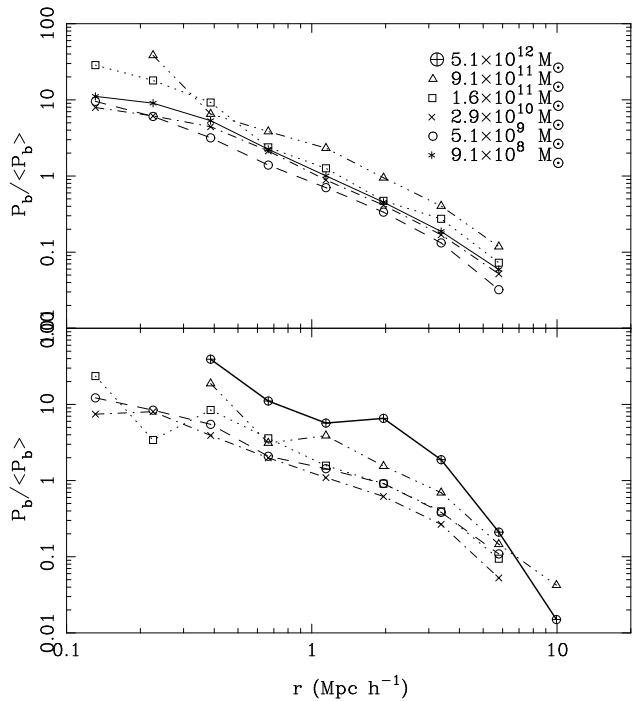}

\caption{Baryonic density in 
units of the cosmic mean as a function of $r$ (distance
from a dark matter halo) in real space for the D5 
simulation(top) and
  G6 simulation 
(bottom). We show results for different dark matter halo masses.
} 
\label{fig:bden}
\end{figure}

Because the density profile around quasars in the simulation is directly
available to us from the simulation data, we examine this first, before
moving on to study the \lya\ absorption profile.
In order to compute the baryonic density profile, we loop over all quasars
in a mass bin and all pixels in the spectra. For each quasar-pixel pair
we compute the separation, $r$ in comoving $\hmpc$. We average the
density values for all pairs at a given separation and show the
results in Figure~\ref{fig:bden}. 

We naturally 
expect that the baryonic density should be a decreasing function of r 
and that the
overall amplitude should monotonically increase as the quasar host mass
increases (Figure~\ref{fig:bden}.) 
We find that this is generally the case, except for the fact that the 
 the smallest mass bin in each simulation 
shows a higher density profile amplitude than expected (we will see later that
this also the case for the mean flux
profile). When we look at the overlap region for the two simulations we
will see later that this is clearly a resolution effect. For the bins
in quasar host mass which are well resolved in the two simulations there is
good agreement.


%

\begin{figure}
\psfig{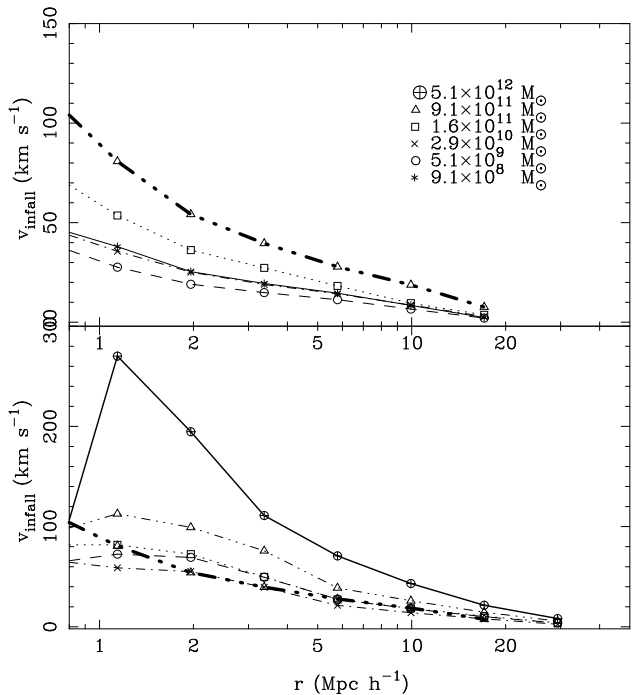}
\caption{Infall velocity as a function of $r$ in real space for 
the simulations D5 (left) and 
  G6 (right). Note that the plot scales are different because of different
  simulation  size. The $v_{\rm infall}$ curve for the largest  mass bin
  $\left<M\right>=2.9\times10^{10}$ in 
the D5 simulation is shown as  bold lines with crosses.} 
\label{fig:vinfall}
\end{figure}

We also compute the mean infall velocity as a function of quasar-pixel 
separation.
The infall velocity ($v_{\rm infall}$)is defined as the
velocity component pointing
towards the 
center of a quasar host. It should depend on 
the gravitational pull of the mass surrounding a quasar  
and therefore decrease as $r$ does and increase as quasar host halo
 mass increases
(Figure~\ref{fig:vinfall}). The tangential component will 
add up to zero if we average over the velocity of gas
around a quasar.
As seen in the case of the baryonic density,
the resolution effect may be responsible for some mass bins having 
higher $v_{infall}$ than predicted. There is discrepancy between 
the two runs for the same mass bin on large $r$ scales. This is
likely to be because the box size of the D5 run is much smaller and
peculiar velocities are sensitive to large-scale density
fluctuations. 

\subsection{Mean \lya\ forest flux as a function of
quasar-pixel distance}

For each mass bin, we calculate the mean absorbed flux defined as:
\begin{equation}
\left<F(r)\right> = \sum \exp^{-\tau(r)},
\end{equation}
where $\tau(r)$ is the optical depth for \lya\
absorption in redshift space, in a pixel at comoving distance from the
quasar $r$. 
The observed flux is the fraction of photons with a given 
wavelength that are left unabsorbed by neutral hydrogen.
Since the \lya absorption
should reflect the presence of neutral hydrogen, we expect the observed flux
to be correlated with the baryonic density, which increases close to 
quasars (small quasar-pixel separation $r$) 
as shown in the previous section.

\begin{figure}
\psfig{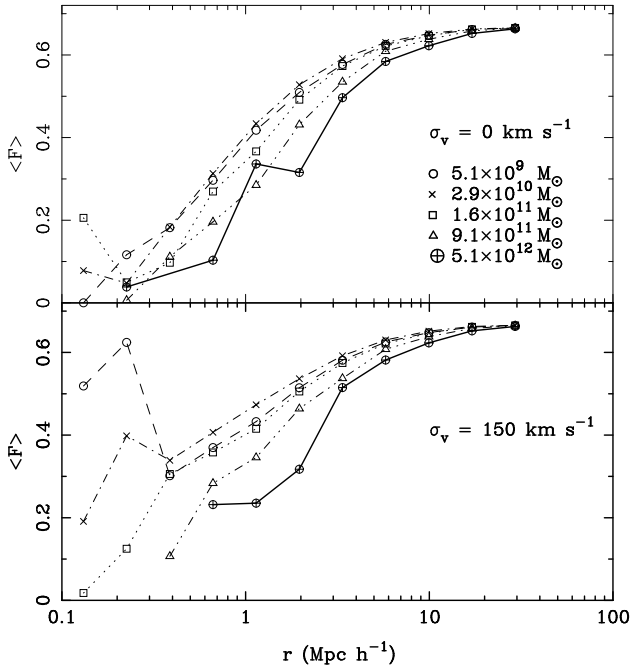}
\caption{The mean transmitted  flux in the \lya\ forest in redshift space,
as a function of quasar-pixel distance. Top: no redshift error has been 
  added. Bottom: a Gaussian redshift error of 150 $\kms$ has been added. The
  flux asymptotically approaches $\left<F\right>$=0.67, the mean
 value that has
  been set in the simulation. 
}
\label{fig:meanflux2}
\end{figure}

We use will use the G6 run, averaging over 500 spectra in this section 
 as its mass resolution is suitable for the halo
 mass range relevant for quasars.
We plot the mean flux in 
 Figure~\ref{fig:meanflux2} where we see  that the
 flux asymptotically
approaches a mean value on large scales, which is 0.67 in this 
case, the value set in the simulation. 
The bottom panel of Figure~\ref{fig:meanflux2} shows the case when we add a
Gaussian redshift error of 150 $\kms$ to the quasar position. This is
to reflect the measurement errors on observational determination of
the quasar redshift.
We can see that in this case, the 
shape of F(r) is different on 
small scales, $r< 1\hmpc$ and also scatters more,
especially for smaller mass halos.
 Since this is the G6 run, it may
not be accurately reflecting the behaviour of quasar hosts of small mass
in any case.
 On larger scales, which will be
relevant for our fitting later, there is not much change.

 The plot shows that the overall flux level $F(r)$
decreases with mass, because more baryonic matter near QSOs means  more
absorptions of photons by the \lya\ forest, which results in 
less observed flux. As with the 
baryonic density (Figure~\ref{fig:bden}) , 
we see that the smallest mean flux is not from the smallest
mass bin. However, we expect this to be due to resolution effects. 
Apart from this, there is a good correlation between absorption and halo mass.

\subsection{Mean flux as a function of quasar-pixel 
separation perpendicular and 
parallel to the line of sight}

\begin{figure}

\psfig{file=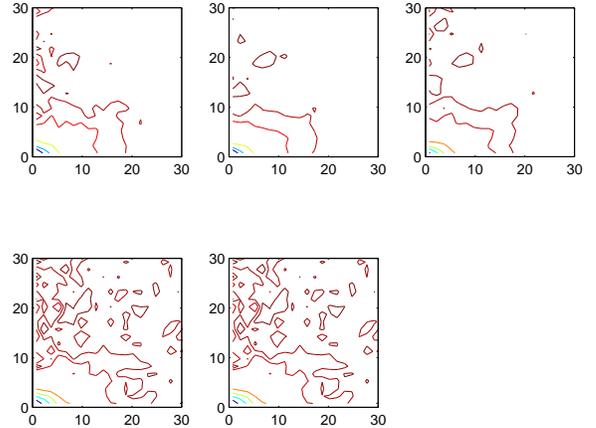,width=9.3truecm}
\caption{The mean observed flux in terms of transeverse and line of sight
  directions. The flux goes through redshift distortions due to peculiar
  velocities and looks squashed in the LOS direction. From the top left:
  $5.1\times 10^{9}~ \msun,  2.9\times 10^{10}~ \msun, 1.6\times 10^{11}~ \msun,
  9.1\times 10^{11}~ \msun$ and $5.1\times 10^{12~} \msun$.  
}
\label{fig:psflux}
\end{figure}

The flux in redshift space is subject to redshift distortions because of
peculiar velocities, which act in a way similar to their effect on the 
 autocorrelation
function (Kaiser 1987).
The baryons around a quasar will infall towards it
because of the gravity from the overdense region. This will 
 result in the compression  of the  flux profile in the line of
sight direction. The squashing effect will be greater for quasars with of 
larger host mass as they are associated with larger density fluctuations. 
We decompose the quasar-pixel separation $r$ into the transverse($\sigma$) and 
line of sight ($\pi$) directions ($r^{2} = \sigma^{2} +\pi^{2}$) and plot the
flux as a function of these two parameters in Figure~\ref{fig:psflux}. 
Although its dependence on quasar mass is not very obvious, the 
distortion effect  is clearly seen. 
In \S5, we measure the anisotropy of the flux from simulations 
in redshift
space by using Hamilton's formulae (1992) to quantify it and 
compare it with observational
data.

\section{Fit to the  Observed Flux vs quasar-pixel separation trend}

In this section  we  parameterise $F(r)$ using
 $r_{0}$ and $\gamma$ which are
commonly used in the power-law auto correlation function regularly
applied to galaxy clustering data. Our aim is to eventually
 estimate the mass of SDSS quasar host halos and of Lyman Break Galaxy
halos by 
using the relation between the fitted values of $r_{0}$ and halo masses. 

We use 5000 spectra from the G6 run,
computing the mean flux as a function of $r$ for 5 different 
mass bins. 
To fit the $F(r)$ we need to use a covariance matrix, as the bins
are correlated. 
In order to make a covariance matrix, we
divide the simulation box into 125 subvolumes. We calculate a full covariance
matrix ($\sigma_{ij}^{M}$) from the mass bin that has most 
quasars. For the rest of mass bins there are not enough quasars to
compute a covariance matrix that is noiseless enough to invert.
Instead we infer the full covariance matrix by scaling up from the matrix
computed from large number of quasars, making use of
 the diagonal elements, which can be
calculated. We assume that the relative cross-correlations
are the same for each covariance matrix. We therefore
only calculate the diagonal elements ($\sigma_{ii}^{m}$) directly and 
calculate the off-diagonal elements from $\sigma_{ij}^{M}$ in the following way:
\begin{equation}
\sigma_{ij}^{m}=\sigma_{ij}^{M} \frac{\sqrt{\sigma_{ii}^{m}}\sqrt{\sigma_{jj}^{m}}}{\sqrt{\sigma_{ii}^{M}}\sqrt{\sigma_{jj}^{M}}}.
\label{eqn:cov_mat}
\end{equation}
We use the mean flux as a function of quasar-pixel difference
computed for 5 different mass bins. We then make curves for 
a total of 80 mass bins (18 bins
in-between each) by linear interpolation. We also build a covariance matrix
accordingly for each sub-mass bin, so that we can compute a smooth trend 
of fit parameters with mass.

\subsection{F(r) fit as a function of halo mass}
In order to find an equation that best describes $F(r)$, we try a power law
form that is commonly used to fit the two point correlation 
function:
\begin{equation}
F(r) = 
\exp\left[-\tau_{\rm eff}\left(1+\left(\frac{r}{r_{0}}\right)^{-\gamma}\right)\right]
\label{eqn:theeqn}
\end{equation}
This formula describes the mean flux as an increasing function with $r$ that 
asymptotically approaches a certain value (because there will be no
relative enhancement of neutral
hydrogen far from a quasar). As $r$
becomes large, the second term in the exponent will decrease, leaving the
the flux close to $\exp^{-\tau_{\rm eff}}$ which is the asymptotic value, and
when $r$ r is small, \lya absorption  depends strongly on the distribution of
matter around a quasar.
\begin{figure}
\psfig{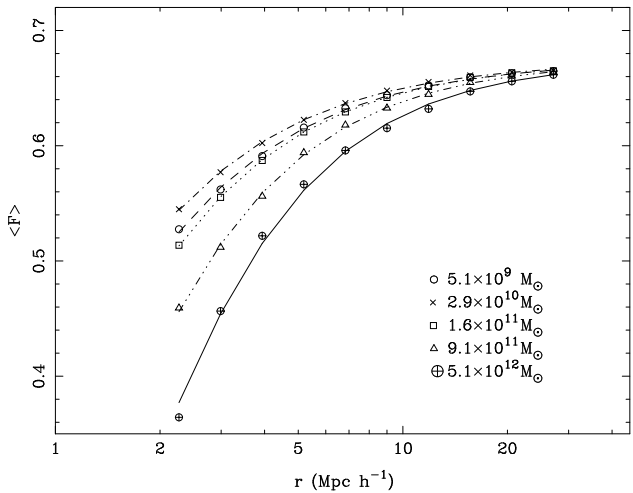}
\caption{The mean flux as a function of quasar-pixel distance
 from the G6 simulation (symbols) and fits using 
  equation~\ref{eqn:theeqn} (lines.) }
\label{fig:frfit_s5k}
\end{figure}

The value of \taueff is set so that the
asymptotic value of equation~\ref{eqn:theeqn} matches the one from the
simulation. We fit the observed flux from the G6 simulation to
equation~\ref{eqn:theeqn} and
show the results in Figure~\ref{fig:frfit_s5k}.
 These results show that even though it is 
very simple,  equation~\ref{eqn:theeqn}
gives a very good description of the observed mean flux
as a function of scale, at least for the
$r$ range of interest.

\begin{figure}
\psfig{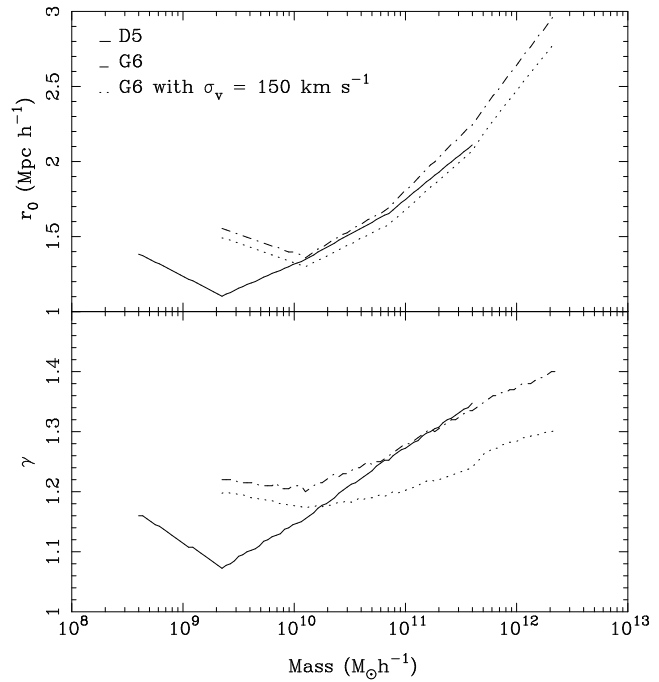}
\caption{The best fit $r_{0}$ and $\gamma$ fitted to the
quasar host-\lya\ forest profile vs. quasar host halo mass.}
\label{fig:r0gm_d5g6}
\end{figure}

We have fitted $r_{0}$ and $\gamma$ for the 80 mass bins,
carrying out
 a  $\chi^{2}$ analysis using the full covariance matrix built as explained
previously. Figure~\ref{fig:r0gm_d5g6} shows the best fit values of $r_{0}$ and
$\gamma$ for the D5 and G6 simulations as a function 
quasar mass. Both $r_{0}$ and
$\gamma$ decrease at first and then go up, which is consistent with the
non-monotonic behaviour of the mean flux in terms of mass as shown in
Figure~\ref{fig:meanflux2}. As we mentioned before, this is due to 
resolution effects, as we can see by for example examining what happens
to halos which approach the resolution limit of the G6 simulation, and 
compare them to the well resolved D5 halos of the same mass.

We see that the slope is getting steeper when
quasar mass increases, which implies that 
the overall amplitude of the flux on small
scales decreases faster.
Since the G6 run is more suited to the mass range of quasar hosts,
we use the $r_0$-mass trends from this run in our fitting , although
our conclusions are not sensitive to this.

We also carry out the same fitting procedure with the G6 run,
after a Gaussian
redshift error with $\sigma=150 \kms$
 is added to the quasar redshifts. Both $r_{0}$ and 
$\gamma$ decrease with the addition of the
redshift errors (also shown in Figure~\ref{fig:r0gm_d5g6}).
The $r_{0}$ value is not much affected by the redshift errors, however.

%
%
%
\subsection{Application to observational data}

Our next step is use the $r_{0}$-mass relation to estimate the mass of
dark matter halos for 
quasars observed in the SDSS, and LBGs, both at redshift $z=3$.
 We will use equation~\ref{eqn:theeqn} to find the
$r_{0}$ and $\gamma$ values that best describe the mean flux 
as a function of separation and then
use the best fit $r_{0}$ to estimate the mass. 

\section{Estimated Mass Limit for SDSS quasars}

We use $\sim$3000 observed quasars from SDSS data release
3 (Abazajian \etal 2005), 
the subset of the total data that have redshifts $z>2.4$. This results in a
mean redshift for the \lya\ forest pixels used in our analysis of $z=3.0$,
matched to the simulations we have been using. This quasar sample
is rather bright, with mean $G$ magnitude  -27.5. 
 It would be ideal to have many
pairs of close-by quasars but 
unfortunately they are rare in observations. Of the $3000$ quasars,
only $\sim 100$ are close enough to contribute to the $F(r)$ for $r<20 \hmpc$.
In order to compute $F(r)$, the pixels in the SDSS spectra we convert the
pixel redshifts and quasar positions into comoving cartesian coordinates
assuming a $\Lambda$ CDM cosmology and then compute quasar-pixel
separations. Only foreground quasars are used in our analysis, so that 
we are not sensitive to the regular proximity effect.
The procedure is the same as in Croft (2004), where it is
outlined in more detail.

\begin{figure}
\psfig{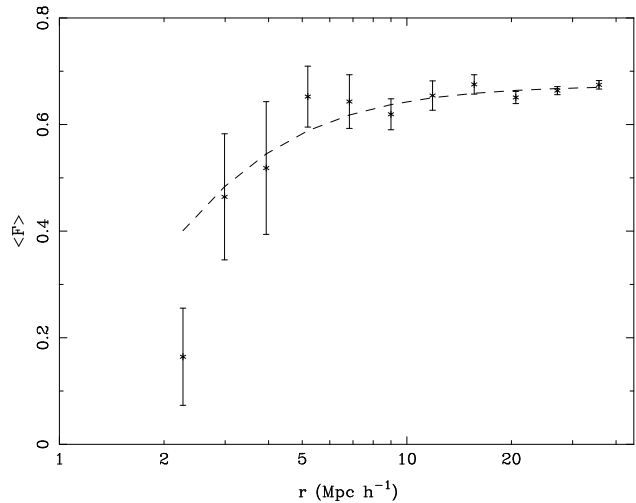}
\caption{The mean transmitted flux 
as a  function of quasar-pixel distance, for SDSS DR3 quasars 
(crosses with error bars) fitted with
  Eqn.~\ref{eqn:theeqn} (dotted lines).}
\label{fig:r0gm_sloan}
\end{figure}

The results for $F(r)$ for the SDSS quasars are shown in Figure 
\ref{fig:r0gm_sloan}, along with the best fit power law curve and jackknife
error bars. In order
to carry out the fit, we use the covariance matrix derived from the
simulations, scaled so that the diagonal elements are the same as the 
observational error bars.

Using the $r_{0}$ vs. mass trend for the G6 simulation, we find the 
estimated mass for SDSS quasars host dark matter halos is
$\log {\rm M/\msun}$=$12.34^{+0.41}_{-0.73}$. Although not used in 
the analysis directly,
fitting $\gamma$ helps narrow down the quasar mass. We also 
tried fixing a $\gamma$
(not necessarily the best fit value) and varied only $r_{0}$ but could not
recover a mass limit. 
In the case of the Gaussian error of 150  $\kms$, added to the
quasar redshifts, the estimated mass limit
becomes 
$\log {\rm M/\msun}$=$12.48^{+0.41}_{-0.73}$. The result agrees 
 well with that 
when the redshift error is not included, making the best fit mass slightly
larger, by $\sim 20\%$. Given the large error bars on the halo mass, this is 
not significant.

The details of quasar formation and evolution in time are still in debate,
which makes it difficult to predict the mass of quasar hosts, and there are
few estimates  their these masses  from observational data. 
Wyithe \& Padmanabhan (2006) report that mass estimates of quasar host
dark matter haloes are from $10^{11} \msun - 10^{13} \msun $, 
depending on the time-dependence of their evolution. 
Our value lies within this range of results.

\subsection{Estimated Mass Limit for Lyman Break Galaxies}

We apply our method to Lyman Break Galaxy data (Adelberger \etal
2003). The reported mass of LBGs from 
other methods at $z=2.9$ is $10^{11.2}-10^{11.8}$
(Adelberger \etal 2005) while other groups have slight different estimates:
$10^{11.6\pm0.3}\msun$ (Weatherley \& Warren 2005) and $10^{11.3}\msun$
(Somerville \etal 2001).

\begin{figure}
\psfig{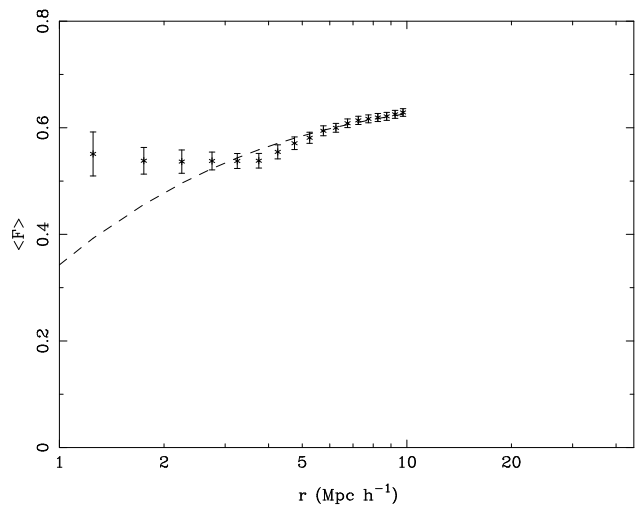}
\caption{The mean flux of Adelberger's data (crosses with error bars) fitted
  to Eqn.~\ref{eqn:theeqn} (dotted lines)} 
\label{fig:r0gm_adel}
\end{figure}

The fitting procedure is same as previously used except that we ignore the 
the first three bins because the data for 
small $r$ is not consistent with the trend in the
simulation despite the large
 error bars(Figure~\ref{fig:r0gm_adel}). If the data points are valid, it would
need further investigation, and indeed these results may be indicative
of starburst winds disturbing the \lya\ forest on small scales.
 But for 
simplicity, and because winds from galaxies are unlikely to propagate far,
we only use regions greater than 1.5$\hmpc$ in our analysis.
We fit the LBG data to the equation~\ref{eqn:theeqn} and then estimate the mass
using the $r_{0}$-mass relation found in \S3.1. 
The estimated mass of LBGs we find is 
 $\log {\rm M/\msun}$=$11.13^{+0.18}_{-0.23}$.
The best fit value of mass is consistent with Adelberger's estimate.


\subsection{Dependence on cosmological parameters}
\begin{figure}
\psfig{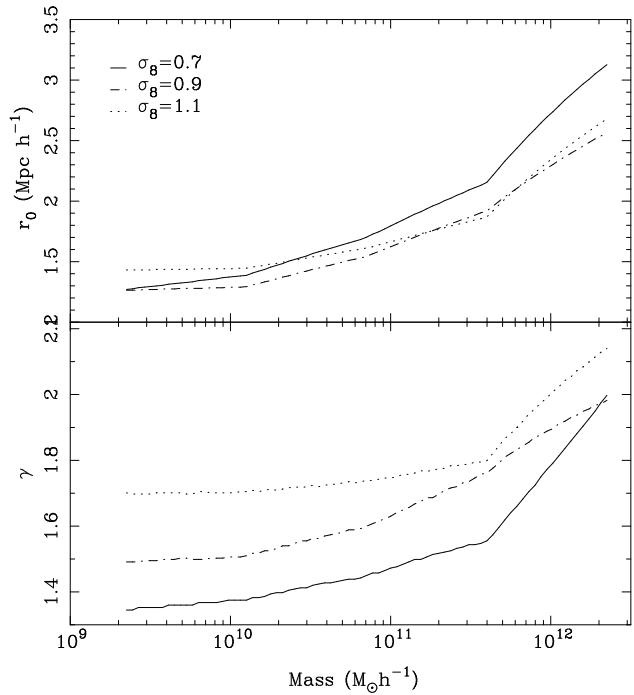}
\caption{$r_{0}$ and $\gamma$ vs. quasar mass for different $\sigma_{8}$ 
  values.(0.7, 0.9, and 1.1) No hydrodynamics is included in the
  simulation. No redshift error has been added. }
\label{fig:r0gm_dm}
\end{figure}
In carrying out our analysis and computing constraints on the halo
masses of quasars and LBGs we have assumed that the $r_{0}$- mass
relation from the simulations applies to the observations.
This is probably a valid assumption, as the cosmological model
used in our simulation is close to that found from for example the 
WMAP satellite data (Spergel \etal 2003, 2006). As the error bars on our
best fit quasar halo mass are large, we expect 
errors on cosmological parameters to have relatively little effect.
 In order to verify
this, we vary the least constrained parameters, the amplitude of
mass fluctuations, $\sigma_{8}$, running new simulations with 
different amplitudes and seeing if we get the same quasar host
masses if we use these to compute the $r_{0}$-mass relation. 

The new simulations are of dark matter only,
and it is assumed that the gas traces the dark matter
distribution. Three runs are made, with  $\sigma_8$=0.7, 0.9 and 1.1.
 They are run in a 50 $\hmpc$ box, using $256^{3}$
 particles. Spectra are made from the dark matter distributions
in the manner described by e.g. Croft \etal (1998). Although the simulations
do not include any hydrodynamics we expect the relative values of 
 $r_{0}$  for the different $\sigma_{8}$ runs to closely approximate the
values that would be obtained with full hydro simulations. 

 The fitting procedure is the same as the D5/G6 simulation cases except
that the covariance matrix was built from 27 subvolumes because the simulation
size is smaller than  the G6 run. We fit the quasar data from SDSS using
equation~\ref{eqn:theeqn} and find the mass limits using the newly found
$r_{0}$ vs. mass relation.  We use no redshift error in the analysis. We find
the mass limits 
as follows:
 $\log {\rm M/\msun}$=$12.21^{+0.36}_{-0.54}$,
 $\log {\rm M/\msun}$=$12.88^{+0.48}_{-0.91}$,
and $\log {\rm M/\msun}$=$12.68^{+0.41}_{-0.75}$ for
$\sigma_8$ = 0.7, 0.9, and 1.1 respectively. The comparison with the results in
\S3.1 ($\log {\rm M/\msun}$=$12.34^{+0.41}_{-0.73}$) indicates that the mass
fluctuations do not have a significant effect on estimating quasar masses,
at least within our error bars (less than a $1 \sigma$ effect).

\subsection{Dependence on \taueff and temperature}
\begin{figure}
\psfig{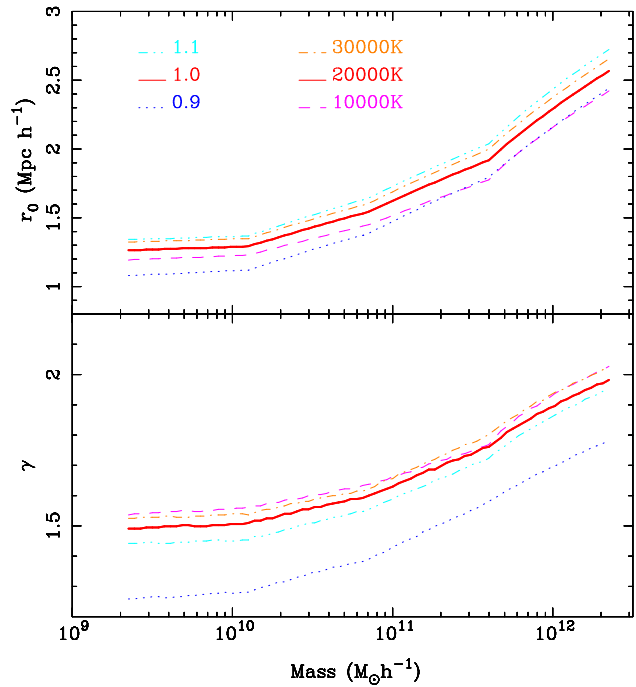}
\caption{$r_{0}$ and $\gamma$ vs. quasar mass when mean observed fluxes are
  90 and 110 \% of the nominal value (0.67) and the gas temperature is 10,000K
  and 30,000K. The solid bold lines represent $\left<F(r)\right>$ =0.67 and
  T=20,000K case ($\sigma_{8}$=0.9) No hydrodynamics is included in the
  simulation. No redshift error has been added. } 
\label{fig:rgm_all}
\end{figure}
The hydrodynamic simulations have a gas temperature at the mean density 
close to 20,000 K (see observational determinations by e.g.
Schaye \etal 2000)
and mean effective
optical depth ($\tau_{\rm eff}=-\ln \left< F \right>$)=
0.4, which has 
been set to be consistent with observational data.
Both of these quantities have uncertainties, and these may affect our
recovery of the quasar dark matter halo mass from \lya\ forest clustering.
 We use the dark matter only simulation to estimate what the total
errors on halo mass will be by quantifying the effect of changes in 
these parameters. We make \lya\ spectra from the simulation 
 setting $T_{0} = 10,000$ K and 30,000 K instead of 
20,000K. We also make spectra by changing
the  the mean
flux $\left< F \right>$ by $\pm10\%$ from 
our original value of   $\left< F \right>=0.67$
In these cases the mean transmitted flux becomes 0.603 and 0.737,
which translates to mean effective optical
depths \taueff of 0.305 and 0.506, respectively. This
range of \taueff  covers the values measured by \taueff by Rauch \etal
(1997) and other observed values:
For example McDonald \etal (2000) reported 
\taueff $0.380 \pm 9$\% at z=3 and Schaye \etal (2003)
obtained $\left< F \right>=0.696$ at z=3.
Although we vary $\left< F \right>$ by $\pm10\%$, this range is much larger
than the scatter in the observed values, which we conservatively
estimate to be $\pm5\%$ at the 1 $\sigma$ level.
We use this value when we compute the effect of $\left< F \right>$ uncertainty
on the quasar host halo mass.

The fitting procedure is the same as in the previous
sections. Figure~\ref{fig:rgm_all} shows the mass vs. $r_{0}$ relation for
each of these cases. 
Adding the uncertainty in quadrature for asymmetrical error bars, we find:
$\sqrt{0.285^{2} + 0.41^{2} + 0.19^{2}} = 0.53$ for the upper error bar and
$\sqrt{0.49^{2} + 0.73^{2} + 0.11^{2}} = 0.89$ for the lower error bar,
 where each term
represents errors from the mean flux, statistical errors (estimated using the
G6 run), and temperature uncertainty
respectively. Including  the 150 $\kms$ velocity uncertainty,
 our mass estimate for $z=3$ SDSS quasar
host halos  becomes $\log {\rm M/\msun}$=$12.48^{+0.53}_{-0.89}$.
This value is in good agreement with the reported mass of quasar host dark 
halo ($M_{DMH}$)   $M_{DMH} = 3.0\pm 1.6\times10^{12}\hmsun$
inferred by Croom \etal 2005 from the autocorrelation function of
quasars in 2dF data.
It is also within 1 $\sigma$ of  $M_{DMH} = 5.2\pm 0.6\times10^{12}\hmpc$ 
recently obtained by Myers
\etal (2006) from clustering of 300,000 photometrically 
classified SDSS QSO. In the same manner, when we add the $\left< F \right>$
errors and temperature errors to the LBG mass estimate, we
obtain $\log {\rm M/\msun}$=$11.13^{+0.39}_{-0.55}$. 

\section{Measuring the Redshift Space Anisotropy of Clustering}

\subsection{Quadrupole to Monopole Ratio}

The galaxy autocorrelation is distorted in redshift space because the line of
sight direction is displaced due to peculiar velocities. 
The overall flow towards
a galaxy mass enhancement will result in a squashed correlation function.
The anisotropy on 
large scales can be parametrized in linear theory using 
$\beta\equiv\Omega_{m}^{0.6}/b$ where $b$ is the
bias parameter (Kaiser 1987, Hamilton 1992.) 
Hamilton (1992) derived the 
expected value of the quadrupole to monopole ratio of the
correlation function 
in terms of $\beta$, using linear theory.
Even on non-linear scales, this ratio can be
used as a measure of the squashing effect.
This measure has been employed by both the SDSS and 2dF survey groups 
to quantify the 
 anisotropy of galaxy clustering in redshift space. (e.g., Zehavi
\etal 2002, Hawkins \etal 2003) 
Replacing the correlation function $\xi(r)$ in Hamilton (1992) with our
quasar-\lya\ forest cross-correlation function, 
$F(r) \equiv F(r)/\left<F\right>-1$ where $\left<F\right> = 0.67$,
we will compute the quadrupole to monopole ratio and see
if it has a dependence on  quasar host halo mass.

Following Hamilton's formulation, we decompose the mean flux
$F(r)$ into $F(r,\mu)$ using Legendre Polynomials:
\begin{equation}
F(r,\mu) = \sum_{l}F_{l}(r)P_{l}(\mu),
\end{equation}
where $\mu$ is the cosine of the angle between the line of sight and the pair
separation vector in redshift space, and $P_{l}$ are Legendre
Polynomials. Integrating over all cosine angles, we get:
\begin{equation}
F_{l} = \int_{0}^{1}F(r,\mu)(1+2l))P_{l}(\mu)d\mu,
\end{equation}
We calculate the quadrupole moment of the flux as follows (Hamilton 1992):
\begin{eqnarray}
Q(s) & = & \frac{(4/3)\beta+(4/7)\beta^{2}}{1+(2/3)\beta+(1/5)\beta^{2}} \nonumber\\
     & = & \frac{F_{2}(s)}{(3/s^{3})\int^{s}_{0} F_{0}(s')s'^{2}ds'-F_{0}(s)}
\end{eqnarray}
We carry this out for the G6 simulation and also for the SDSS quasar
data. Our expectation is that the magnitude of quadrupole moment is 
greater for more massive quasar host halos. If this
is the case, the goal is then to use the relation
between $M_{DMH}$ and the quadrupole moment to constrain the mass of 
SDSS quasars in a manner independent of the overall strength of absorption
which we have use in the main part of this paper.

\begin{figure}
\psfig{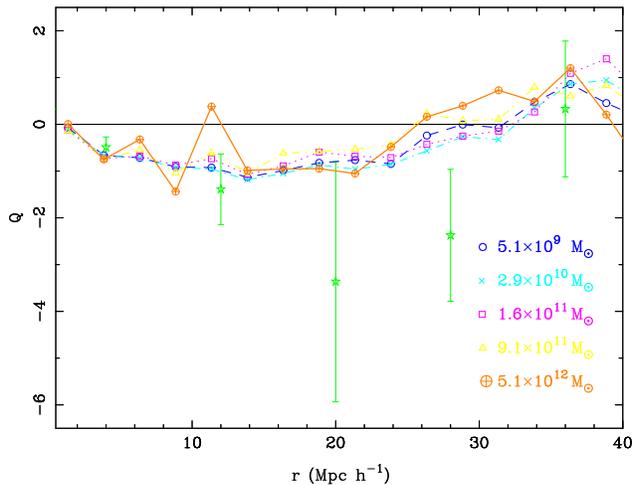}
\caption{Quadrupole moments of the mean flux from the G6 simulation and SDSS
  data (stars with error bars). No redshift error has been
  added.} 
\label{fig:qm}
\end{figure}

In  Figure~\ref{fig:qm} we show the quadrupole to monopole ratio for 
halos in several mass bins.
Contrary to our expectation, the quadrupole moment does not show a very strong
dependence on $M_{DMH}$ (Figure~\ref{fig:qm}). It is not
possible to see any
useful monotonic relation between Q(s) and $M_{DMH}$.

We have computed Q(s) from our SDSS quasar sample, and find
results that are consistent with the simulations, but with 
extremely large error bars. From this study, it seems as though
the simulations reproduce the anisotropy of clustering, but there is
not any expectation of being able to use its magnitude to measure quasar 
host halo mass. On the other hand, the anisotropy will be 
sensitive to assumed cosmology, and a much
larger sample of this type of data could be used to carry out an 
Alcock-Paczy\'{n}ski (1979) test.

\section{Summary and Discussion}

\subsection{Summary}

In this paper, we have explored the profile of \lya\ absorption
with distance around quasars using pairs
of close-by quasars in high resolution hydrodynamic simulations. We have found
the following:

\begin{itemize}
\item The mean \lya\ forest transmitted flux from a background quasar passing 
through a  region close to a foreground
  quasar and the infall velocity of matter around a quasar are both
 dependent on quasar mass.

\item We can fit the observed flux vs quasar-pixel separation $r$
to an equation $\exp(-\tau_{\rm eff}
  (1+(\frac{r}{r_{0}})^{-\gamma}))$ by varying $r_{0}$ and $\gamma$.
This simple power law form, derived from fits to
the galaxy autocorrelation function works very well.

\item  There is a strong dependence of $r_{0}$ in this fit to the
mass of the quasar host halo in the simulation.
From this $r_{0}$ vs. mass relation, we can estimate quasar halo mass.

\item The estimated mass of SDSS quasars at $z=3$ is $\log {\rm
  M/\msun}$=$12.48^{+0.53}_{-0.89}$. 
We do not see any significant changes in the result
  when we include redshift errors of ~150$\kms$ or a different amplitude 
of mass fluctuations.

\item The estimated halo  mass of Lyman Break Galaxies 
can be found using the same technique. Fitting to the data of
Adelberger (2003) we find  $\log {\rm M/\msun} =
  11.13^{+0.39}_{-0.55}$.
 
\item The squashing of flux in redshift space can be
quantified using the quadrupole
  to monopole ratio.  There is a no measureable dependence
of the squashing effect on halo mass, although the 
anisotropy seen is consistent with  observational data.

\end{itemize}

\subsection{Discussion}

Due to their great distance from us, it is difficult to
measure the dark matter halo masses of high redshift 
quasars. For example, few background 
galaxies exist to make lensing measurements possible, and the rotation
curves of their host galaxies may require larger telescopes than currently
available to capture. 

The strong level of \lya\ forest absorption around bright $z=3$ 
SDSS quasars that we have seen
indictates that they have halo masses $\sim 3\times10^{12} \msun$,
comparable to elliptical galaxies at the present. This is not unreasonable
because these bright quasars are much rarer than present
day ellipticals with this mass. Our mass estimate is roughly 20 times larger
than that for the dark matter mass of LBGs at the same redshift, indicating
that it is unlikely that bright quasars randomly sample the population
of LBGs as hosts.

In our analysis, we have ignored the transverse proximity effect,
the possiblity that foreground quasars will decrease the amount of \lya\ 
absorption with local photoionization from their radiation. If this effect
is present, it is certainly not dominant, as we have seen that instead there
is more absorption close to foreground quasars. If it does exist, but is
subdominant, then the true amount of absorption without this extra radiation
should be greater and we are underestimating it and hence the halo masses.
 $\sim 3\times10^{12} \msun$ therefore represents a lower limit to the 
dark matter mass of quasar hosts. There are however several ways, including
quasar beaming and short lifetimes mentioned in \S1 that the transverse
proximity effect might not be relevant in any case.

The redshift distortion of the absorption is unfortunately
not promising as a complimentary method for constraining
the dark matter halo mass. However the insensitity of the distortion to
mass means that the correlation function it could be a useful probe  
of cosmic geometry. As the \lya\ forest- quasar clustering signal 
appears to be much stronger \lya\ forest two point clustering,
 this may be an efficient way of carrying out a high-z Alcock Paczy\'{n}ski test.

\section*{Acknowledgments}
We thank Scott Burles for providing us with 
the SDSS DR3 \lya\ forest sample used here, and also 
Kurt Adelberger for providing us his LBG data in machine readable form. 
We also thank Volker Springel and Lars Hernquist
for allowing us to use the cosmological
simulation data.

\end{document}